\documentclass[a4paper,fleqn,usenatbib]{mnras}
\pdfoutput=1

\usepackage{newtxtext,newtxmath}

\usepackage[T1]{fontenc}
\usepackage{ae,aecompl}

\usepackage{graphicx}	
\usepackage{amsmath}	
\usepackage{amssymb}	

\title[Suppressed Central SF in Seyferts]{Mildly Suppressed Star Formation in Central Regions of
  MaNGA Seyfert Galaxies}

\author[L.Bing et al.]{
Longji Bing,$^{1,2}$ 
Yong Shi,$^{1,2}$ \thanks{E-mail: yong@nju.edu.cn} 
Yanmei Chen,$^{1,2}$ 
Sebasti\'an F. S\'anchez,$^{3}$ 
\newauthor
Roberto Maiolino,$^{4}$ 
Rog\'erio Riffel,$^{5,7}$ 
Rogemar A. Riffel,$^{6,7}$ 
\newauthor
Dominika Wylezalek,$^{8}$
Dmitry Bizyaev,$^{9,10,11}$ 
Kaike Pan, $^{9}$ 
and Niv Drory$^{12}$
\\
$^{1}$School of Astronomy and Space Science, Nanjing University, Nanjing 210093, China.\\
$^{2}$Key Laboratory of Modern Astronomy and Astrophysics (Nanjing University).\\
$^{3}$Instituto de Astronomia, Universidad Nacional Autonoma de Mexico, A.P. 70-264, 04510, Mexico, D.F., Mexico.\\
$^{4}$Cavendish Laboratory, University of Cambridge, 19 J. J. Thomson Avenue, Cambridge CB3 0HE, United Kingdom.\\
$^{5}$Instituto de F\'isica, Universidade Federal do Rio Grande do Sul, Campus do Vale, Porto Alegre, RS, Brasil, 91501-970.\\
$^{6}$Laborat\'orio Interinstitucional de e-Astronomia, 77 Rua General Jos\'e Cristino, Rio de Janeiro, 20921-400, Brasil.\\
$^{7}$Departamento de F\'isica, CCNE, Universidade Federal de Santa Maria, 97105-900, Santa Maria, RS, Brazil.\\
$^{8}$European Southern Observatory, Garching, Germany.\\
$^{9}$Apache Point Observatory and New Mexico State University, P.O. Box 59, Sunspot, NM, 88349-0059, USA.\\
$^{10}$Sternberg Astronomical Institute, Moscow State University, Moscow.\\
$^{11}$Special Astrophysical Observatory of the Russian AS, 369167, Nizhnij Arkhyz, Russia.\\
$^{12}$McDonald Observatory, The University of Texas at Austin, 1 University Station, Austin, TX 78712, USA.
}

\date{Accepted XXX. Received YYY; in original form ZZZ}

\pubyear{2018}
\hypersetup{draft}
\begin{document}
\label{firstpage}
\pagerange{\pageref{firstpage}--\pageref{lastpage}}
\maketitle
\begin{abstract}
Negative feedback from accretion onto super-massive black holes (SMBHs), that is to remove gas and suppress star formation in
galaxies, has been widely suggested.  However, for Seyfert galaxies which harbor less active, moderately accreting SMBHs in the local universe, 
the feedback capability of their black hole activity is elusive. We present spatially-resolved H$\alpha$ measurements  to
trace ongoing  star formation in Seyfert  galaxies and compare their specific star formation rate with a sample of star-forming galaxies whose global 
galaxy properties are controlled to be the same as the Seyferts. From the comparison we find that the star formation rates within central kpc
of Seyfert galaxies are mildly suppressed as compared to the matched normal star forming galaxies. This suggests that the feedback of moderate 
SMBH accretion could, to some extent, regulate the ongoing star formation in these intermediate to late type galaxies under secular evolution. 

\end{abstract}




\section{Introduction}

Super-massive black-holes (SMBHs) reside at centers of almost all
massive galaxies\citep{Kormendy13}.The accretion onto SMBHs,
which powers phenomena known as active galactic nuclei(AGN), is suggested to play
key roles in driving galaxy evolution by depositing accretion energy
into the interstellar medium(ISM) in AGN host galaxies to regulate
star formation\citep{King03, Springel05, Hopkins06}. The feedback
could be in forms of removing gas from the central part of host
galaxies by fast multiphase outflows\citep{Maiolino12, Cano-Diaz12, 
Cicone14, Carniani16}, or heating the gas within and surrounding the
host galaxies by radio jets\citep{Forman07, Fabian12}. Observations have 
found the former case in luminous quasars associated with major mergers, while 
evidence for the latter case mainly comes from radio-loud AGNs reside in massive red elliptical galaxies.

Although AGNs with moderate SMBH accretion (Seyfert galaxies) are much
more numerous, their capability of feedback is unclear. It is still
unknown if  the SMBH's  feedback in  Seyferts is necessary in
re-producing the SMBH-bulge relationships because bulges form through
the major merging while Seyfert galaxies are mostly blue and disk dominated spirals in secular evolution
\citep{Hopkins06, Kormendy13, Heckman14}. This is in contrast to the
case for massive red radio galaxies with elliptical morphology and
 the remnants  of major  merging\citep{Springel05,  Cheung16}, and
luminous quasars   that  are  associated   with   ongoing  major   merging
\citep{Springel05, Hopkins06}.  Although outflows of ionized gas 
are seen in Seyfert galaxies,  the observed low outflow rates indicate
that they could be driven by the nuclear star formation itself\citep{Harrison14, Ho14, Wild14, Lopez-Coba17a}.
Some recent spatially-resolved studies of a handful  nearby Seyferts
reveal  the  existence  of  fast   outflows  of  ionized or dense  molecular  gas
associated with  radio jets but their impacts  on star  formation are
unclear\citep{Christensen06, Krause07, Wang12, Garcia-Burillo14, Morganti15, Querejeta16, Lopez-Coba17b}.

Observationally it is a big challenge to conclude whether AGN's
feedback could  regulate star formation or not, and how it regulates
star formation. On one hand, if strong outflows emerge, they could
clear out the star-forming  gas to suppress star formation
\citep{Alexander12, Garcia-Burillo14, Alatalo15, Hopkins16, Wylezalek16q}.  
The heating by jets  propagating through the
galaxies could prevent gas from cooling  and cutoff the gas supply for
further star formation\citep{Karouzos14, Choi15}. On the other hand,
the outflows and  jets interact  with the  gas in host galaxies  and
compress it  to  trigger new star formation \citep{Zubovas13, 
Silk13, Zubovas17}. In fact observations show either no or positive
relationships between star formation  rates and SMBH  accretion rates
but no negative trends are seen \citep{Shi07, Shi09, Baum10, Xu15, Zhang16, Mallmann18}.
Whether AGN's feedback plays the role may also depends on
the spatial scale that observations could resolve and timescale that the 
observed tracers could probe\citep{Harrison12, Feruglio15,
  Cresci15_blow}. For example, radiation from AGNs nearly
instantaneously impact the surrounding  ISM while the attenuation from
ISM probably limits their impact to the nuclear regions
\citep{Roos15}. Outflow or jets travels slowly and may be decelerated
after interactions with ambient gas, which delays their  effects on
star formation at large distances from  the nuclei \citep{Harrison17,
  Harrison18}. Feedback by jets or outflows on ISM also depends on
their orientation relative to the dusty torus, making their effects on
star formation  to be  anisotropic. The short  duty cycle of 
AGNs  could  also make  feedback  by  radiation from AGNs temporally
variable in strength. Case studies of individual AGNs find evidence of
coexistence of positive and negative feedback on star
formation\citep{Zinn13}, suggesting the complicated nature of AGN
feedback.

The availability of spatially-resolved spectra as enabled by integral
field unit (IFU) observations offers a new opportunity to investigate
the possible effects of AGN's feedback on star formation. These data
allow measurements of a range of host galaxy properties in details
including the star formation rate, stellar mass, stellar population,
and metallicity at kpc scales or smaller and relate them to the AGN activities
\citep{Davies07,Dumas07, Riffel11, Wylezalek17, Riffel17, Sanchez18}.  Especially, a large IFU-observed sample
of nearby  galaxies from on-going SDSS-IV MaNGA  \citep{Bundy15} offers
the opportunity to statistically explore relationships between AGN and
star formation at kpc scales. In this study we present studies of
spatially-resolved star formation activities of nearby Seyfert
galaxies and comparisons with a control sample of normal star-forming
galaxies.  Only with the large number of galaxies observed by MaNGA can we construct such a control sample
that all physical parameters that may affect star formation are
controlled. In Section 2, we describe the basic  information of MaNGA
data, the selection of AGN and control samples and the procedures of 
measuring the intensity of spatial resolved star formation. The difference 
in spatially-resolved SFRs between MaNGA AGNs and their comparison 
samples of normal galaxies are shown in section 3. We discuss the possible 
implications of our results in the context of galaxy evolution in section 4.  
A flat $\Lambda$CDM cosmology with $\Omega_\Lambda$=0.692, $\Omega_M$=0.308 and $H_0$=67.8 km/s/Mpc is assumed throughout this study.

\section{Data and Methods.}

\subsection{MaNGA Data}

We analyzed  a large sample of 4756 galaxies with spatially-resolved
integral field unit (IFU) observations carried out by the program of
Mapping Nearby Galaxies at Apache Point Observatory (MaNGA)
\citep{Bundy15, Yan16}. MaNGA is an ongoing integral field unit(IFU) survey to acquire
spatially resolved spectra of nearby $\sim$10000 galaxies from 2014 to
2020 \citep{Bundy15, Drory15, Law15, Yan16, Yan16_cal, Blanton17}, with the Sloan Digital Sky
Survey (SDSS) 2.5m telescope  \citep{Gunn06} and the Baryon Oscillation
Spectroscopic  Survey   (BOSS)  spectrograph\citep{Smee12, Drory15}.  MaNGA
galaxies have a flat distribution in the stellar mass above 10$^{9}$
M$_{\odot}$ and are composed of two sub-samples with different radial
coverage: a primary sub-sample of about 70\% galaxies ($<z>$=0.03)
with a coverage to 1.5 effective radii($R_{\rm e}$) and a secondary sub-sample of about
30\% galaxies ($<z>$=0.04) with a coverage to 2.5 $R_{\rm  e}$. The
spectrum covers a wavelength range from 3600 to 10300 \AA\ with a
velocity resolution $\sigma$ of $\sim$ 65 km/s. The spatial
resolution is about 2.5''. The S/N at $r$ band is about 4-8 \AA\ $^{-1}$
at the edge of the radial coverage.

The raw data were reduced, calibrated and reconstructed to a data
cube  by the  Data Reduction  Pipeline (DRP) \citep{Law16}. The {\sc Pipe3D}
\citep{Sanchez16a,Sanchez16b} was  applied to the data cube to measure
the continuum and associated physical qualities. Both {\sc Pipe3D} and MaNGA Data Analysis Pipeline(DAP) (Westfall et al. in prep)
provide measurements on emission line fluxes that we used in sample selection and
star formation rate calculation. These measurements  were available in
the internal MaNGA Product Launch currently at version 6 (MPL\_6). These IFU spectra
will be released in the coming SDSS DR15.

\subsection{The selection of AGN with star-forming disks}

\begin{figure*}
  \includegraphics[scale=0.9]{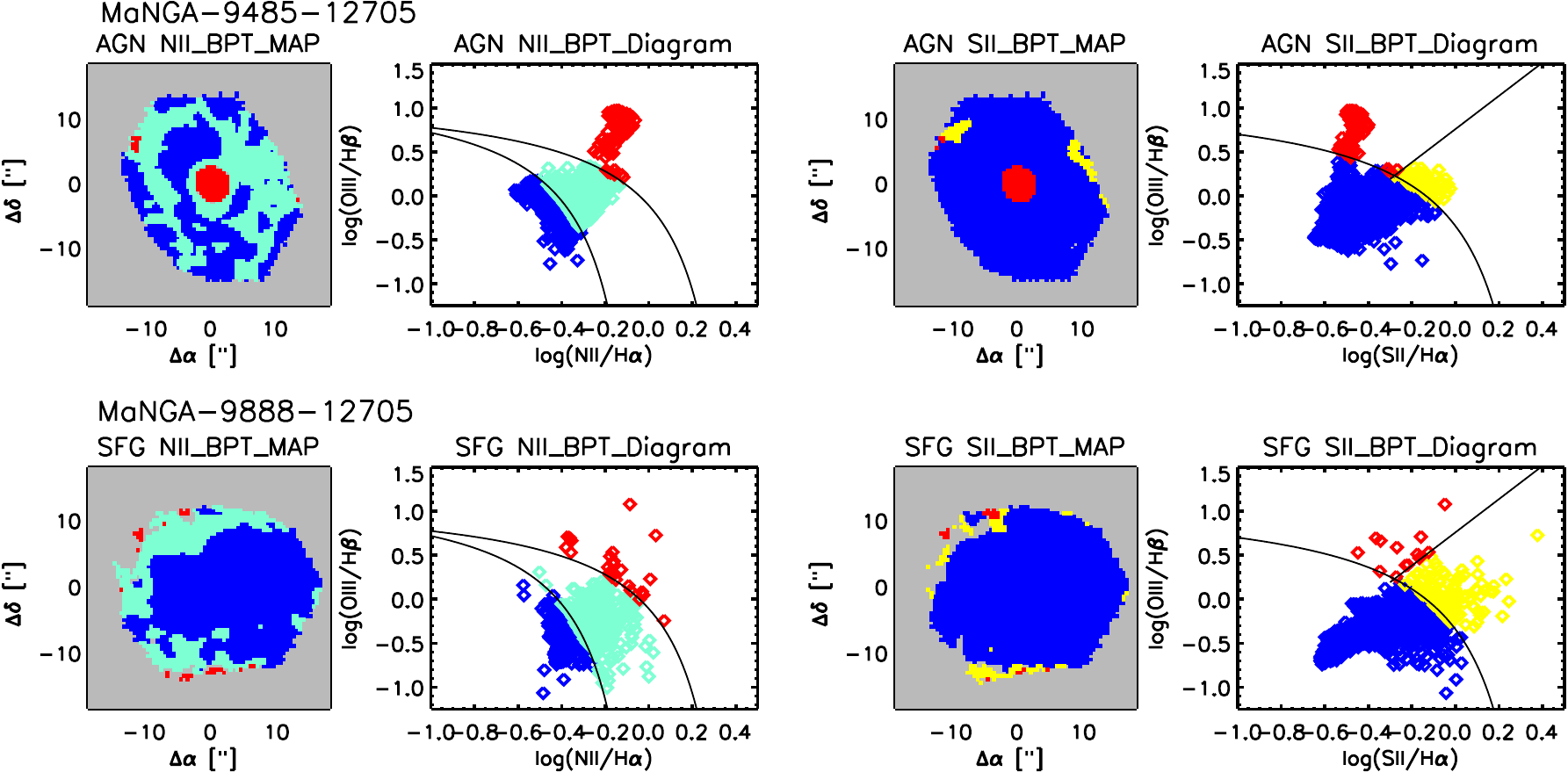}
  \caption{Example BPT maps  and diagrams of galaxies  selected as our
    AGN samples and  control samples. Ionization in different part of
    galaxies are classified by BPT diagram, where we mark AGN/Seyfert
    region in red, LINER region in yellow (only [SII]/H$\alpha$ BPT
    diagram), composite region in cyan (only [NII]/H$\alpha$ BPT
    diagram) and star forming region in blue in resolved BPT map. The
    lines in BPT diagrams shows the diagnostic criteria from
    \citet{Kauffmann03} and \citet{Kewley06}.}
    \label{fig:bpt_exp}
\end{figure*}

For each MaNGA galaxy, we first constructed the spatially-resolved BPT
diagram (both NII-BPT and SII-BPT) to classify them into star-forming,
composite and AGN (also LINER) according  to the dividing lines in the
literature  \citep{BPT81, Kauffmann03,  Kewley01, Kewley06}. We used
all spaxels  with S/N  of H$\alpha$, H$\beta$,  [OIII] 5007\AA\ $ $ and
[NII] 6584\AA\  (or [SII] 6717/6731\AA\ ) larger than 2 measured by MaNGA DAP 
using the pure emission line spectrum. A galaxy
is classified as an AGN as long as the emission from the central spaxel
of  the  MaNGA  datacube   has  [NII]/H$\alpha$,  [SII]/H$\alpha$  and
[OIII]/H$\beta$ line ratios that satisfy: 1. either the composite or the
AGN definition in the [NII]/H$\alpha$  BPT diagram. 2. the AGN
or  LINER  definition in  the  [SII]/H$\alpha$  BPT-diagram.  We  also
required the equivalent width of H$\alpha$ emission($EW_{H\alpha}$) at
the central spaxel to be larger than 3\AA\ ,  which rejects galaxies
with weak emission lines powered by evolved stars \citep{Stasinska08, Cid-Fernandes10, Yan12,
 Belfiore16}. Spaxels classified as 'star formation' are further
required to have $EW_{H\alpha} > 6$ \AA\ as suggested by previous
studies \citep{Sanchez14}. We then selected AGN with
star-forming disks as those objects with AGN present at galaxy centers
and star formation in the outer regions to investigate the effects of
AGN's  feedback on star formation. Quantitatively, we required that more
than half of spaxels within 1.0-1.5 $R_{\rm e}$ should be filled by
star formation spaxels, as we need the sSFR at this radial bin to be properly measured to constrain the 
same parameter in control galaxies to be the same(see section 2.4 for details). Edge-on galaxies with minor/major axis ratio in NSA catalog $<$
0.35 \citep{Blanton11} are excluded to eliminate the additional smearing of the light
from adjacent radial bins at high inclinations.  A total of 56 AGNs are
selected under these criteria and an example is shown in Fig.~\ref{fig:bpt_exp}.

\subsection{The selection bias of the AGN sample}

Because the  line emission at the  center are contributed not  only by
the AGN narrow line region but also by star formation, the BPT diagram
misses AGN  with high central  SFRs.  We  performed a test  to qualify
this bias  in our  sample selection, following  the spirit  of related
methods  applied  in  previous  studies  \citep{Kauffmann09,  Trump15,
  Davies16}.  We generated simulated  AGNs of different  [OIII]
luminosities  by adding PSF-convolved point source  emission to  the
emission line maps  of pure star-forming galaxies. Given  a AGN [OIII]
luminosity, the point  source emission of different  emission lines is
calculated as following:

(1)  For  [OIII] 5007\AA\ $ $  itself,  the  AGN's  contribution  is  from
randomly  selected AGN  $L_{\rm  [OIII]}$ after  the dusty  extinction
based  on the  central Balmer  decrement \citep{Osterbrock06}  and PSF
convolution at $g$-band available in the MaNGA DRP.
 
(2) For  H$\beta$, the AGN's  contribution is calculated by  the above
AGN  [OIII] 5007\AA\  $ $ flux  multiplied with  a [OIII] 5007\AA\ /H$\beta$
flux ratio. We fixed this ratio  to 10 in our simulation, while varying
it to other values does not significantly change our result.

(3) For H$\alpha$,  the AGN's contribution is calculated  by the above
H$\beta$ flux multiplied with the observed Balmer decrement.

(4)  For   [NII] 6584\AA\ $ $ and  [SII] 6717\AA\ +6731\AA\ ,   the  AGN's
contribution is measured  by the above H$\alpha$ flux  multiplied by a
representative [NII] 6584\AA\ /H$\alpha$ or
([SII] 6717\AA\ +6731\AA\ )/H$\alpha$ flux  ratio.  We fixed  these two
ratios to the average values of all selected AGN.

\begin{figure*}
	\includegraphics[scale=0.86]{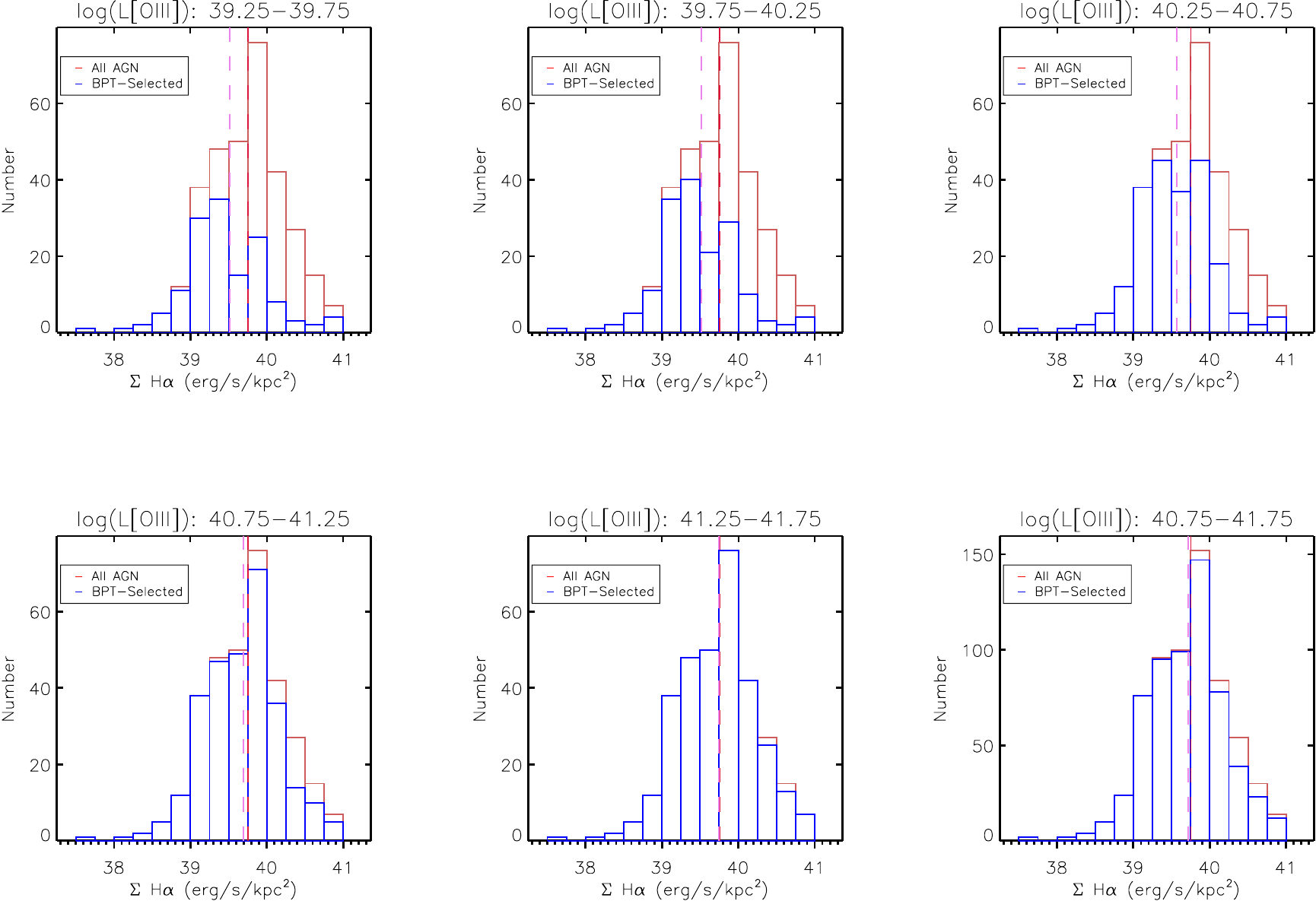}
    \caption{The test on the selection bias of the BPT diagram on central star formation. Blue and red histogram show the distributions of central SFR-related H$\alpha$ surface brightness of all normal galaxies and normal galaxies that are identified as AGN after adding AGN light. Red and pink vertical lines mark the median value of the distributions. Panel 1-5 shows the result of simulation carried out within 5 luminosity bins. Panel 6 shows the combined result of panel 4 and 5, which supports ignorable bias to lower central star formation rate in BPT selected AGNs at $log(L_{[OIII]} > 40.85)$.}
    \label{lumbias}
\end{figure*}

A uniform distribution in the logarithmic [OIII] luminosity is applied
in the  simulation. In total five  luminosity bins were used  to cover
the range from $7\times 10^{39}$ to $7\times10^{41}$ erg/s.  With this
simulated AGN  sample, we  then defined those  that satisfied  the BPT
criteria as the BPT-selected AGN.   Within a [OIII] luminosity bin, we
measured the central $\Sigma_{H\alpha}$ (after attenuation correction)
of the BPT-selected AGN and compared to the central $\Sigma_{H\alpha}$
of all  AGN. Fig.~\ref{lumbias} shows the  distributions of extinction
corrected $\Sigma_{H\alpha}$  of all  simulated AGNs  and BPT-selected
AGNs.    The  figure   clearly  shows   lower  $\Sigma_{H\alpha}$   in
BPT-selected AGNs when the AGN  [OIII] luminosities drop below certain
values. It also  indicates that the above selection  bias becomes less
significant  in  luminous  AGNs   with  the  [OIII]  luminosity  above
7$\times$10$^{40}$ ergs/s.  We thus  restricted our  AGN sample  to be
above this luminosity in order to  avoid the bias toward selecting AGN
with low central SFRs. After applying this luminosity constraint we got a 
final AGN sample with 14 objects in total.

\subsection{The selection of the comparision sample of pure star-forming galaxies for each AGN.}

To  determine  whether the  influence  on  star formation  happens  in
Seyfert  galaxies, a well defined comparison  sample of  pure star-forming  galaxies
without SMBH accretion is needed so that all parameters
that affect  star formation are controlled to be the same as Seyfert
galaxies. We considered six control parameters, including the total stellar mass,
the presence/absence  of the bar structures,  the bulge-to-disk ratio,
the stellar mass surface density at 1.5$R_{e}$, the sSFR at 1.5$R_{e}$
and the central  stellar mass surface density in  defining the control
galaxies  for  each AGN  from  these  pre-selected galaxies  with  the
following quantitative criteria:

(1) The  difference of the total stellar  mass between each AGN  and its
comparison galaxies to  be  within  $\Delta$log($M_{*}$)$<$0.3: the  galaxy
stellar mass controls the overall  global galaxy properties to exclude
any stellar-mass-dependent  effects on  the SFRs \citep{Shi11, Shi18},  such as  the galaxy
metallicity \citep{Shi14} and the  total SFRs \citep{Brinchmann04}.  The stellar
mass  measurement was  taken from  the GALEX-SDSS-WISE  legacy catalog
\citep{Salim16}  based on  the UV-optical  SED fitting.   For galaxies
without  measurement in  \citet{Salim16},  we turned  to stellar  mass
measurement  in   SDSS+WISE  MAGPHYS   output  catalog \citep{Chang15}.
Stellar masses from these two studies are consistent with
each other over the mass range of MaNGA samples(within 0.2dex, see section 8 in \citet{Salim16}).

(2) The strength of the bar structures:  bars could enhance star
formation in  the central regions of galaxies \citep{Kennicutt94}.  We
required the  difference  of  the debiased  vote  fraction  of  ''the
existence  of bar''  from the Galaxy Zoo 2 between  each  AGN and  its
comparison galaxies to be lower than 0.25. \citep{Willett13}.

(3) The difference  in bulge-to-disk ratio between each  AGN and its
comparison  galaxies to  be within  $\Delta B/T  < 0.2$.  Star formation  in
bulges is generally lower than in disks.  The bulge-to-disk ratio thus
relates to  the relative level of  star forming activity in  the inner
parts of galaxies  as compared to the outer  parts.  The bulge-to-disk
ratio  is based  on  the  $r$-band decomposition  of  the SDSS  images
\citep{Simard11}.  For galaxies  without  B/T  measurement in  Simard's
catalog,  we use  the  fracDev  parameter in  $r$  band  from SDSS  DR12
photometric catalog\citep{DR12} as an alternative control parameter on
the bulge-to-disk ratio. Only one of these two parameters is used in the selection 
of comparison galaxies to a specific sample galaxy.

(4-5)  At 1.5  $R_{\rm e}$,  the  difference of the stellar mass  surface
density  between  each  AGN  and  its comparison  galaxies  to  be  within
0.3  dex($\Delta$log$(\Sigma{M_{*,1.5R_e}})$$<$ 0.3  dex) and  the difference  of
sSFR  between  each   AGN  and  its  comparison galaxies   to  be  within 0.3 dex
($\Delta$log(sSFR$_{1.5R_e}$)$<$0.3). These  two constraints  ensure the
AGN and comparison galaxies to have the same levels of disk star formation
as characterized by the SFRs and  sSFRs.  With the  MaNGA-produced H$\alpha$
maps, we can make consistent measurements of the disk SFRs for all objects.  In
addition, by fixing the star formation  level at 1.5 Re, we can assure
any difference  in the  inner region  is due  to the  central activity
instead of  the global offset.  These qualities are measured  by data
products in MaNGA  $MPL\_{6}$.

Here we did not control the global SFRs because: 1. The SFR measurements 
using MaNGA data only are also not consistent because the coverage of IFU bundle to
different galaxies varies\citep{Bundy15, Wake17}. 2. The total SFR is not consistently 
measured for the MaNGA galaxies. The SFR derived by \citet{Salim16} 
and \citet{Chang15} are not well compatible with each other at lower sSFRs.  
3. The inner SFR/sSFR that also contributes
to the total SFR of galaxy is the quality we want to compare between the AGNs and 
comparison galaxies. 

(6) The  difference in  central stellar  mass surface  density between
each    AGN    and   its  comparison    galaxies    to   be    within
$\Delta\Sigma{M_{*,0.3R_{\rm  e}}}$$<$0.5  dex:  this  quantity  could
further constrain the central  structure \citep{Fang13}, and avoid the
effect of any spatially-resolved SFR/stellar-mass relationship on the
central star formation.

\subsection{The Measurements Of Spatially Resolved sSFRs}

With the selection of  AGN with star-forming disks and  comparison star-forming
galaxies, we then carried out measurements of  the  spatial-resolved   specific  star
formation rate (sSFR). Here the specific  star formation rate refers to
the star formation  rate divided by the stellar mass, which qualifies the
current stellar  mass growth  rate. We  used the attenuation corrected
H$\alpha$  flux map  to estimate  the spatial-resolved SFR 
\citep{Kennicutt98}.  The  stellar  mass   surface  density  of  MaNGA
galaxies are measured using {\sc Pipe3D}\citep{Sanchez16a, Sanchez16b}.  
A Salpeter  initial mass  function  is  assumed. Attenuation  correction
applied  to H$\alpha$  flux  assumed the  case B  Balmer
decrement(H$\alpha$/H$\beta$=2.87)   and  the extinction    law   of
\citet{Calzetti01}.

For the  central part of  the AGN where  the emission comes from both
SMBH  accretion  and  star  formation,  we  tried  to  estimate  the
SFRs through decomposition.  The  contribution from star formation in the
Balmer emission line could be derived if the intrinsic [OIII]/H$\beta$ line ratio 
of AGN emission is known. The observed [OIII] 5007\AA\  and H$\beta$ emission
line flux in the central region of these 14 type 2 AGNs could be contributed by both
 star formation and AGN narrow line region, thus the [OIII]/H$\beta$ ratio could be 
 expressed as Eq~\ref{ratio}

\begin{equation}
  \frac{F_{[OIII] 5007}^{obs}}{F_{H\beta}^{obs}} = \frac{F_{[OIII] 5007}^{AGN}+F_{[OIII] 5007}^{SF}}{F_{H\beta}^{AGN}+F_{H\beta}^{SF}}
    \label{ratio}
\end{equation}

The right part of the Eq.\ref{ratio} could also be written in forms of $[OIII]/H\beta$ line ratios of star formation and AGN narrow line region.

\begin{equation}
  \frac{F_{[OIII] 5007}^{obs}}{F_{H\beta}^{obs}} = \frac{F_{[OIII] 5007}^{AGN}}{F_{H\beta}^{AGN}} \times \frac{F_{H\beta}^{AGN}}{F_{H\beta}^{obs}}+\frac{F_{[OIII] 5007}^{SF}}{F_{H\beta}^{SF}} \times \frac{F_{H\beta}^{SF}}{F_{H\beta}^{obs}}
     \label{ratio2}
\end{equation}

The superscripts tell the origin of the emission lines(AGN, star forming region or the observed total emission line flux). The total emission line flux and their line ratio are observables that could be derived directly from MaNGA data while the line ratio at AGN NLR and need to be determined with other information(see the following paragraph and Section 4.1 for the discussion). Then the contribution of star formation in Balmer emission line could be derived based on the observed [OIII]/H$\beta$ line ratio using the Eq~\ref{sffrac} with assumptions on the [OIII]/H$\beta$ line ratio of AGN and star formation base spectra. From Eq~\ref{sffrac} we could express the $\frac{F_{H\beta}^{SF}}{F_{H\beta}^{obs}}$ at the end of Eq~\ref{ratio2} as:

\begin{equation}
    \frac{F_{H\beta}^{SF}}{F_{H\beta}^{obs}} = \frac{({F_{[OIII] 5007}^{AGN}}/{F_{H\beta}^{AGN}})-({F_{[OIII] 5007}^{obs}}/{F_{H\beta}^{obs}})}{({F_{[OIII] 5007}^{AGN}}/{F_{H\beta}^{AGN}})-({F_{[OIII] 5007}^{SF}}/{F_{H\beta}^{SF}})}
	\label{sffrac}
\end{equation}

The [OIII]/H$\beta$ value for  pure star formation in Eq~\ref{sffrac}
are determined for each  galaxy individually, as it is sensitive to the metallicity\citep{Maiolino08}.  We  measured the  average  of
[OIII]/H$\beta$ in different radial bins  with a width of 0.1Re, among
which the lowest  value is used as the ([OIII]/H$\beta$)  of pure star
formation. For the pure AGN line ratio, we applied
log([OIII]/H$\beta$)=1.0 with  a error of $\pm$0.2.  This brackets two
cases: the value of log([OIII]/H$\beta$)=0.8 has previously been used
in  a  similar  analysis  by   \citet{Kauffmann09},  and  a  value  of
log([OIII]/H$\beta$)=1.2 is similar to the value of the upper bound of all 
SDSS AGN and also the maximum of our selected AGNs. The uncertainties of
log([OIII]/H$\beta$) value in AGN region is the major source of error in the 
estimation central SFR($\sim$ 0.2-0.3dex), while the high S/N of MaNGA spectra
makes contribution of measurement error of emission line flux to be minimal in 
this process.

\section{Results}

\subsection{Properties of AGNs and their Host Galaxies}

The general properties of 14 AGNs and their host galaxies are listed in Table~\ref{AGNprop}. We derived the [OIII] luminosity from the reduced MaNGA datacube. 
The lower and upper limit of $L_{[OIII]}$ are calculated from the integrated [OIII] flux within central 2" and that of all regions classified as composite or AGN/LINER 
in BPT diagram. We collected the total SFR and total stellar mass of each galaxy from large multi-wavelength value-added catalogs of \citep{Salim16, Chang15}. 
Our AGN sample have luminosities that are typical of Seyfert galaxies and most of their host galaxies have late type morphology and high stellar masses($log(M/M_{\odot}>10.5)$). We also crossmatched our sample with the FIRST Survey Catalog\citep{Helfand15} to obtain their properties in radio emission. Seven out of fourteen AGNs in our sample have FIRST detection with S/N>5. Three of them have 1.4GHz luminosity higher than the prediction of their total SFRs by more than 0.5dex using converting factor from \citet{Kennicutt12}, which probably suggests the existence of radio jet or outflows\citep{Ho08, Zakamska16,Hwang18}. 

\begin{table*}
	\centering
	\caption{SFR and 1.4GHz luminosity 14 MaNGA AGNs.}
	\label{AGNprop}
	\begin{tabular}{cccccccc} 
		\hline
                 MaNGA Plate-IFU & z & $log(L_{[OIII]})^{a}$  & $log(L_{1.4GHz})^{a}$  & $log(SFR_{1.4GHz})^{b}$ & $log(SFR_{SED})^{b}$ & $log(SFR_{22micron})^{b}$ & $log(M/M_{\odot})^{c}$ \\ 
	         \hline
		$8147-6102$  & 0.0631 & $41.07 \pm 0.14$ & $<29.04$ & $<0.837$                & $0.712 \pm 0.142$    & 0.454 & 11.06 \\
                 $8241-6102$   & 0.0373 & $41.54 \pm 0.11$ & $29.56$  & $1.362 \pm 0.006$  & $0.407 \pm 0.053$    & 1.291 & 10.63 \\
                 $8317-12704$ & 0.0543 & $41.09 \pm 0.12$ & $<28.90$ & $<0.697$                & $-0.011 \pm 0.230$   & N/A   & 11.36 \\
                 $8459-3702^{d,e}$ & 0.0722 & $41.01 \pm 0.15$ & $29.57$  & $1.370 \pm 0.025$  & $0.660 \pm 0.192$    & 0.548 & 11.25 \\
                 $8588-12704$ & 0.0304 & $41.05 \pm 0.13$ & $28.43$  & $0.235 \pm 0.056$  & $-0.111 \pm 0.287$   & N/A   & 10.90 \\
                 $8606-12701$ & 0.0633 & $41.29 \pm 0.13$ & $<29.04$ & $<0.839$                & $0.479 \pm 0.024$    & 0.367 & 11.39 \\
                 $8718-12701$ & 0.0499 & $41.13 \pm 0.12$ & $<28.81$ & $<0.619$                & $0.436 \pm 0.247$    & 0.262 & 10.86 \\
                 $8718-12702^{d}$ & 0.0389 & $41.00 \pm 0.11$ & $29.08$  & $0.881 \pm 0.018$  & $0.146 \pm 0.138$    & 0.377 & 10.71 \\
                 $9047-6104^{d}$ & 0.0586 & $41.46 \pm 0.11$ & $29.81$  & $1.610 \pm 0.009$  & $0.534 \pm 0.164$    & 1.052 & 11.27 \\
                 $9095-12701$ & 0.0875 & $41.22 \pm 0.10$ & $<29.34$ & $<1.144$                & $0.498 \pm 0.244$    & 0.739 & 11.31 \\
                 $9182-12703$ & 0.0688 & $41.24 \pm 0.11$ & $<29.12$ & $<0.916$                & $0.247 \pm 0.133$    & 0.608 & 11.21 \\
                 $9196-12703$ & 0.0819 & $42.00 \pm 0.13$ & $29.59$  & $1.386 \pm 0.031$  & $0.919 \pm 0.118$     & 1.201 & 11.37 \\
                 $9485-12705$ & 0.0323 & $41.35 \pm 0.12$ & $<28.42$ & $<0.222$                & $0.221 \pm 0.081$    & $1.236^{f}$ & 11.05 \\
                 $9508-12704$ & 0.0809 & $41.37 \pm 0.09$ & $29.66$  & $1.455 \pm 0.024$  & $0.192_{-3.615}^{+0.385}$ & N/A   & 11.37 \\
		\hline
		
		\multicolumn{8}{l}{Notes:  $^a$ In $ergs/s$.  $^b$ In $M_{\odot}/yr$.  $^c$ In $M_{\odot}$.  $^d$ Radio luminosity exceeds the prediction of star formation. }\\
		\multicolumn{8}{l}{                           $^e$ Blob source in \citep{Wylezalek17} $^f$ Identified as MIR AGN, $SFR_{SED}$ likely overestimated. }  \\

		\multicolumn{8}{l}{References: 
		(1) This Work;		
		(2) \citet{Salim16};
		(3) \citet{Chang15}; 
		(4) \citet{Helfand15}; 
}\\
	\end{tabular}
\end{table*}

\subsection{Star Formation in AGN Host Galaxies}

\begin{figure*}
  \includegraphics[scale=0.9]{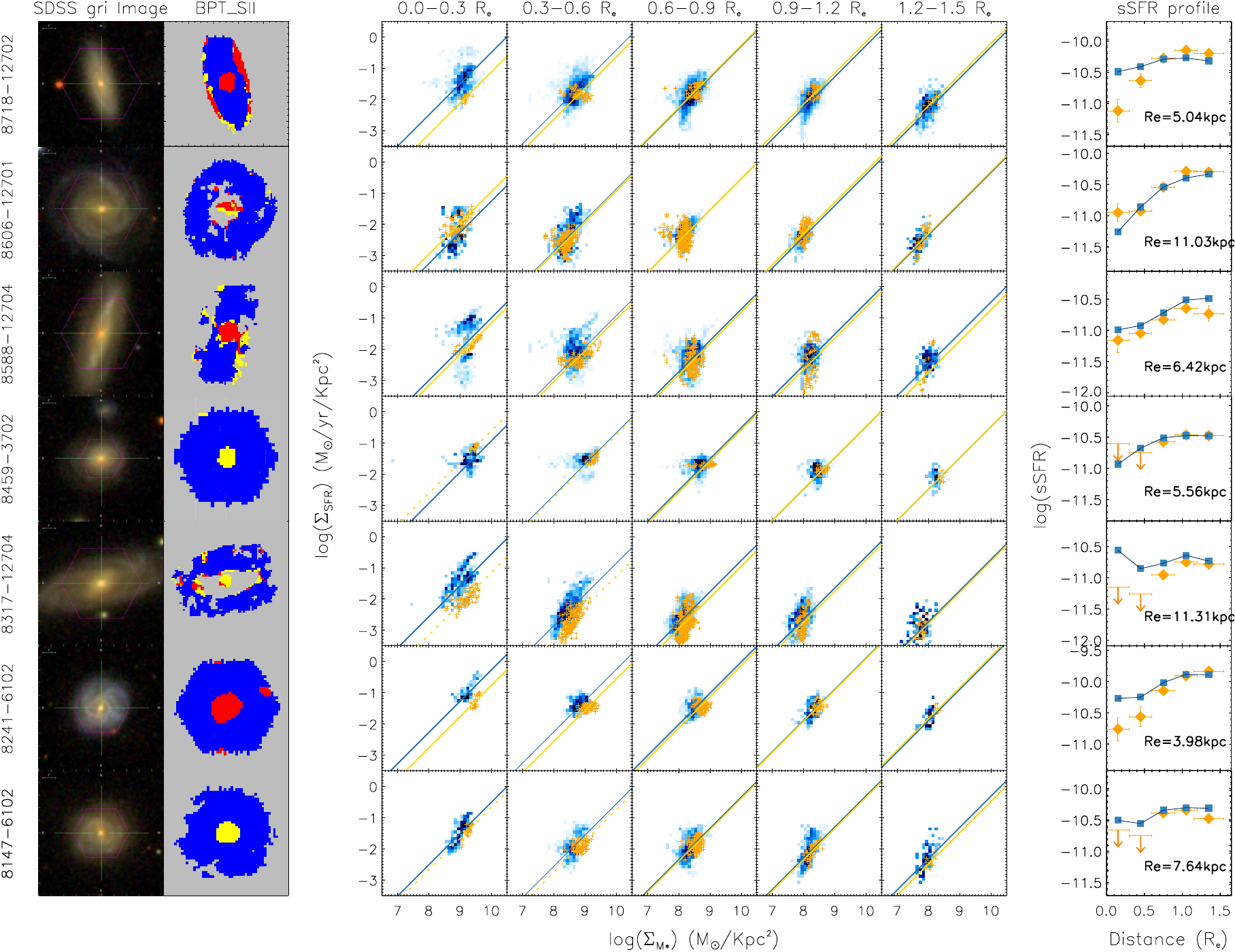}
\caption{ Our selected AGN sample and their star forming properties in 5 radial bins. From left
  to right: the $gri$ false-color images of AGN host galaxies with MaNGA IFU footprint overlaid;
  the BPT maps of AGN host galaxies based on the emission line flux maps produced by {\sc Pipe3D};
  the SFR surface density vs. stellar mass surface density in five radial bins for AGN (yellow points) and their control
  normal galaxies (blue points), where the lines are the best linear fitting to AGN (brown) and
  normal galaxies (pink), respectively; dashed lines are fitting of bins dominate by non-star forming emission in AGN host galaxies; the mean sSFRs (interceptions of the best linear fits) of AGN and normal galaxies
  as a function of radii.}
  \label{first11}
\end{figure*}

\begin{figure*}
  \includegraphics[scale=0.9]{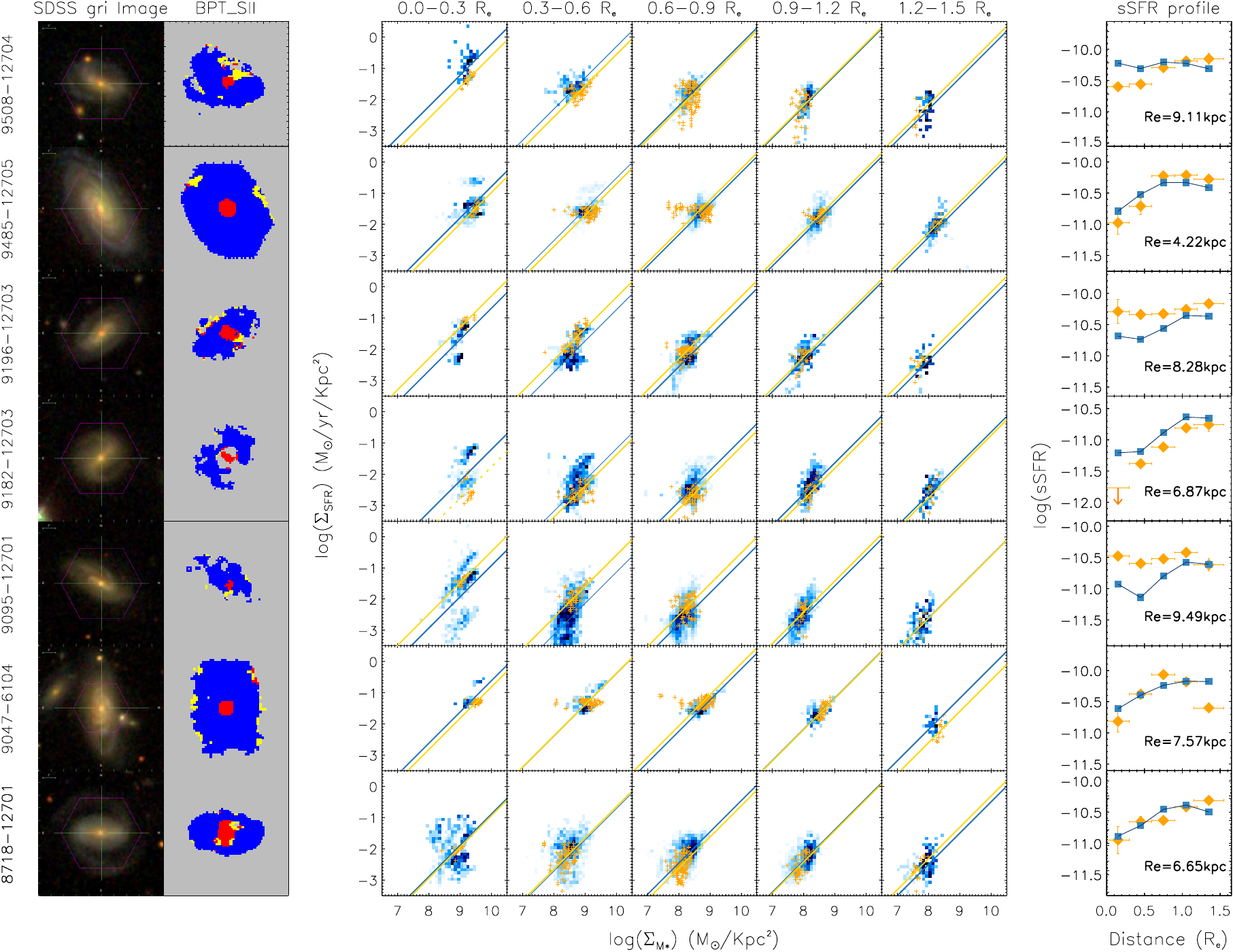}
\caption{ The same as Fig.~\ref{first11} but for additional 7 objects. }
\label{second11}
\end{figure*}

\begin{figure*}
  \includegraphics[scale=0.85]{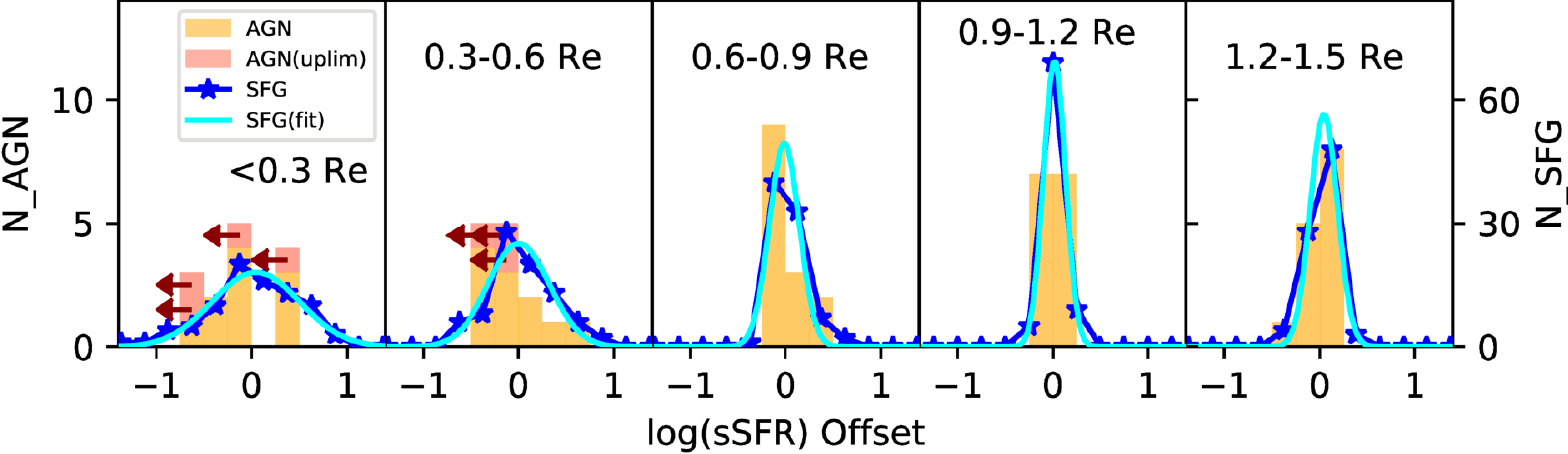}
\caption{ The  distribution of  the difference  in the
  log(sSFR)  between AGN  and their  comparison normal  galaxies.  The
  histograms are for our AGN, the  symbols are the 'fake  AGN'  that are
  randomly retrieved from normal galaxies with matched properties of AGN host galaxies. 
  The solid line is the
  best fitted Gaussian  profile to the fake AGN.  From  left to right,
  the distributions are shown for five radial bins. }
  \label{ssfr_off}
\end{figure*}

\begin{table*}
	\centering
	\caption{Numbers of AGN with Suppressed Star Formation in 2 Central Radial Bins}
	\label{rbin_frac}
	\begin{tabular}{lccr} 
		\hline
		Radial Bin & $\Delta$log(sSFR) $<$ -0.00 & $\Delta$log(sSFR) $<$ -0.25 & $\Delta$log(sSFR) $<$ -0.50 \\
		\hline
		0.0-0.3 $R_{e}$ & $10/14$ & $5/14$ & $3/14$ \\
		0.3-0.6 $R_{e}$ & $10/14$ & $5/14$ & $0/14$ \\
		\hline
	\end{tabular}
\end{table*}

\begin{table*}
	\centering
	\caption{The probability to draw the $\Delta$log(sSFR) distribution of MaNGA Seyferts from star formation galaxies with similar or even lower value.}
	\label{rbin_significance}
	\begin{tabular}{lccc} 
		\hline
		Radial Bin & P($N_{AGN}$) < P($N_{Control}$) \\
		\hline
		0.0-0.3 $R_{e}$ & 0.76\% \\
		0.3-0.6 $R_{e}$ & 3.5\% \\
		\hline
	\end{tabular}
\end{table*}

To  compare the  spatially-resolved  sSFRs between AGNs and  normal galaxies,  
we plotted  the SFR  surface density  vs.  stellar
mass surface density ($\Sigma_{*}$) of  all spaxels within five radial
bins($<$ 0.3  $R_{\rm e}$, 0.3-0.6  $R_{\rm e}$, 0.6-0.9  $R_{\rm e}$,
0.9-1.2 $R_{\rm  e}$ and  1.2-1.5 $R_{\rm  e}$) for  each AGN  and its
comparison sample, respectively. Both quantities are corrected for the
galaxy  inclinations  derived  from  the  photometric  axis  ratio  in
NASA\_Sloan  Atlas \citep{Blanton11}.    The  linear  fitting   to  the
relation between log($\Sigma_{\rm SFR}$) and log($\Sigma_{\rm M_{*}}$)
in  each  radial  bin  was  then performed  for  AGNs  and  comparison
galaxies,  respectively.   With  the  slope  fixed  to  a  unity,  the
interceptions of the linear fitting give the sSFR in each radial bins.
The log(sSFR) as  a function of the galactocentric radii  for each AGN
and their comparison galaxies are shown in the final panel of each row
in the Fig~\ref{first11} and  Fig~\ref{second11}. The relation between
 $\Sigma_{M_*}$ and $\Sigma_{SFR}$ is actually sub-linear as previous studies suggested\citep{Sanchez13, Cano-Diaz16}
However, the selection of the slope does not significantly affect the measured sSFR in each
radial bins. We could derive similar sSFRs as the fitting method even when calculating the mean sSFR in each radial bins. The central sSFRs of
LINERs  are  used  as  upper-limits   given  that  they   could  have
intrinsically lower [OIII]/H$\beta$ values than Seyferts, which causes
an overestimate of their central SFRs in our analysis.

The distributions of the  differences in the sSFRs ($\Delta$log(sSFR))
between AGN and their comparison normal galaxies are shown as
filled histograms  in Fig~\ref{ssfr_off} for  each radial bin.  In the
inner two bins ($<$0.3$R_{\rm e}$ and 0.3-0.6$R_{\rm e}$), the offsets
in sSFRs  are systematically  biased toward negative  values, implying
possible negative feedback.  In outer radial  bins (0.6-0.9, 0.9-1.2
\& 1.2-1.5 $R_{\rm  e}$), $\Delta$log(sSFR) has a  median value around
zero, indicating that the star formation of AGN are not different from
pure star-forming galaxies. As listed in Table~\ref{rbin_frac}, 10 out
of 14  AGN have $\Delta$log(sSFR) $<$  0.0 in the central bin, with 5
AGN below -0.25 and 3 AGN below -0.5.

To compute the statistical significance of these offsets, we 
drew a group of star-forming galaxy with matched physical properties of AGNs
and measured their $\Delta$log(sSFR).  We
used the AGN comparison sample of 75 star-forming galaxies and for
each of them we selected a control sample using the same methods as we
did for the AGN  sample. Then we measured  $\Delta$log(sSFR) of
each star-forming galaxy from its comparison star-formation galaxies.
The star  symbols in Fig~\ref{ssfr_off}  show the results.   We fitted
these distributions with Gaussian  functions to derive the statistical
probability as following:  first to randomly draw 14  objects from the
Gaussian distribution  and then to  compute the probability  that more
objects  than  the observed  number  have  $\Delta$log(sSFR) $<$  0.5,  $\Delta$log(sSFR) $<$  0.25, $\Delta$log(sSFR) $<$  0.0,
$\Delta$log(sSFR) $<$  -0.25 and $\Delta$log(sSFR) $<$  -0.5, e.g. for
the central bin this require all 14 randomly selected galaxies with $\Delta$log(sSFR) $<$
0.5, $>$ 10 galaxies with $\Delta$log(sSFR) $<$
0.0, $>$ 10 AGN with $\Delta$log(sSFR) $<$
0.0, $>$  5 AGN with  $\Delta$log(sSFR) $<$ -0.25  and $>$ 3  AGN with
$\Delta$log(sSFR)       $<$       -0.5.       As       listed       in
Table~\ref{rbin_significance}, the probability to produce the observed
$\Delta$log(sSFR)  distribution  of  the   MaNGA  Seyferts  from  star
formation galaxies (P($N_{AGN}$) < P($N_{SF}$)) is as low as 0.76\% in the innermost bin and 3.5\% in the
second radial bin, supporting that the central regions of AGN most likely
have suppressed SFRs.

\section{Discussion}

\subsection{The Emission Line Flux Decomposition}

Some assumptions are made about the decomposition of the central emission
line  fluxes. We consider the observed fluxes of central emission lines are the
superposition of the  emission from massive young stars  and that from
SMBH  accretion. Under such assumption, a similar method  has  been applied to  high
spatial-resolution  IFU spectra of some nearby
galaxies \citep{Davies14,Davies16}. They  decomposed the  emission line
by  using the  intrinsic  spectra of  different ionization  mechanisms
(star formation, AGN or shock). However, the poor spatial resolution of
MaNGA data  (about factor of 5  larger in the real  physical scale per
spaxel)  causes  significant  blending  between  the   emission  from
star-formation regions and AGN NLR regions,  making it hard to find an
central spaxel with purely star-formation region or AGN NLR emission.
This dilution
could  be supported  by the  much lower  maximum [OIII]/H$\beta$  line
ratio in some  MaNGA AGNs  compared  to that in
high-spatial resolution  IFU data\citep{Davies16} and  to the largest values of all
SDSS   samples (e.g.  \citet{Brinchmann04}).  To overcome caveats due to our low spatial resolutions,
we adopted fixed maximal line ratios of AGN, which, if any, causes the derived SFRs to be overestimated, thus
strengthening our conclusions of suppresses central SFRs in AGN hosts. 

Our decomposition method also assumes the same Balmer decrement for AGN NLRs narrow
as star-forming HII  regions. Previous studies of local AGNs have found that AGN NLRs
do have  similar dust extinctions to  the  HII regions in their  host galaxies \citep{Wild11,Trump15}, which
holds over a large range of obscuration.  Since  our AGN  samples  are  also selected  optically,  it
is  unlikely that  they have different internal dust extinction properties compare to AGNs  in
\citet{Wild11, Trump15}. Significantly larger extinction in single HII
region has been seen in the ENLR  of Centaurs A
from high resolution observations  \citep{Salome16}. However, the derived young age makes
this HII region an extreme  among extragalactic HII regions. And the poor spatial resolution of the
MaNGA fiber makes the observation only sensitive to kpc-scale average extinction, which is similar to studies of
 \citet{Wild11, Trump15}. If we consider the higher intrinsic Balmer decrement in AGN dominated regions\citep{Osterbrock06}, 
the central star formation rate corrected by the case B value could be slightly overestimated, while this does not dramatically 
change the main results.

\begin{figure}
  \includegraphics[width=\columnwidth]{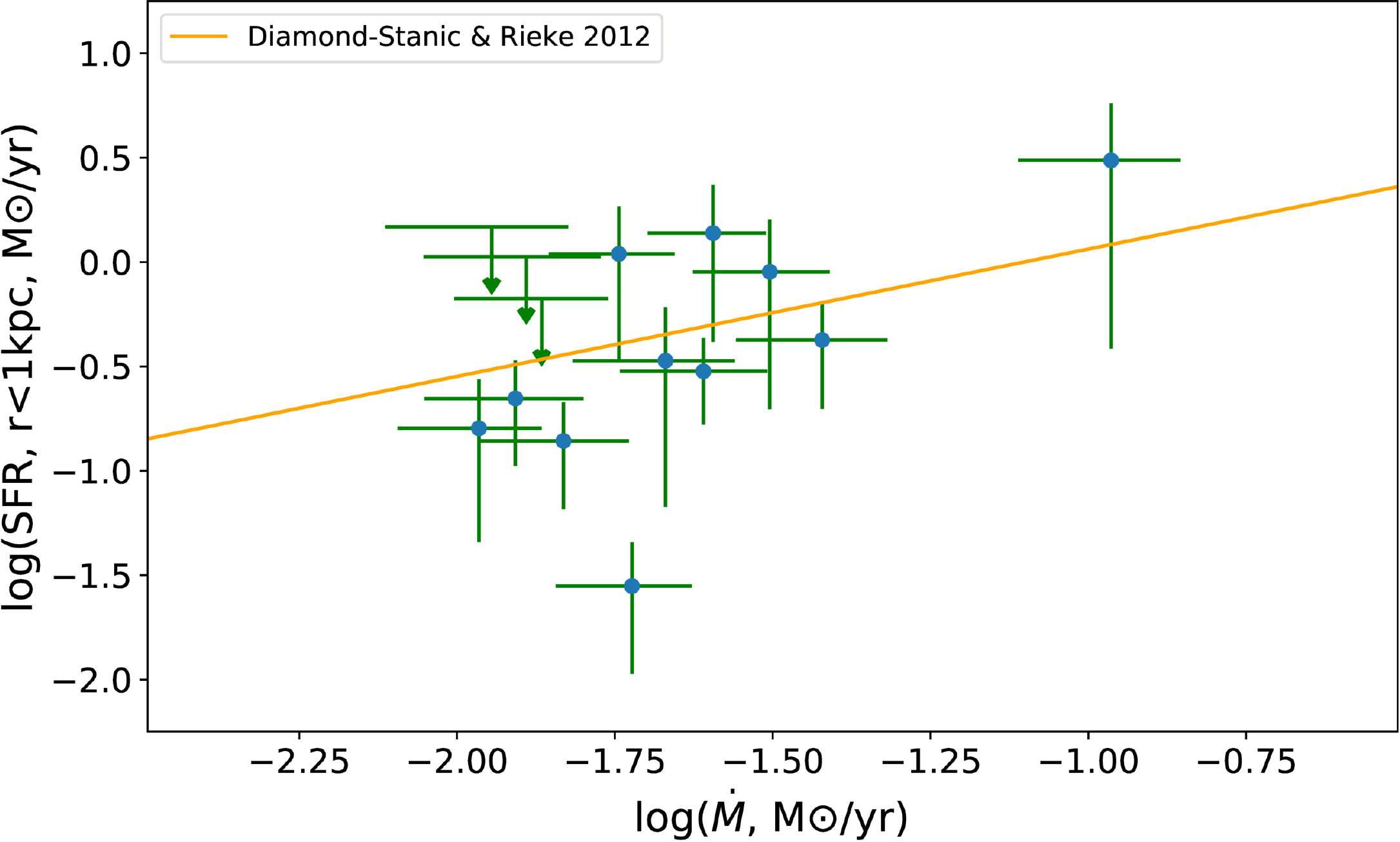}
  \caption{The distribution of our AGN on the plot of the nuclear SFR vs BHAR where the solid
    line is the best-fitting one in the literature \citep{Diamond-Stanic12}. }
    \label{nuclearSFR_BHAR} 
\end{figure}

We examined the  reliability of  our SFR  measurements of  our AGN
through  the  relation  between  central  SFR and black hole accretion rate(BHAR)  as  found  in
\citet{Diamond-Stanic12}.    They   derived   the   SFRs   using   the
mid-infrared  aromatic  features  that  are insensitive  to  the  dust
extinction and contamination by  AGN. We used our decomposition
method to derive the SFR within the same aperture  ($< 1kpc$)  as in
\citet{Diamond-Stanic12}  and  the  BHAR from  the  [OIII]  luminosity
\citep{Kauffmann09}  for all our 14 AGNs.  As shown in
Fig.~\ref{nuclearSFR_BHAR},  our derived  SFRs  and  BHARs in  general
follows the relation as found by \citet{Diamond-Stanic12}, suggesting
the reliability of our SFR measurements.

\subsection{Implications of AGN feedback on Star Formation}

\begin{figure*}
  \includegraphics[scale=0.8]{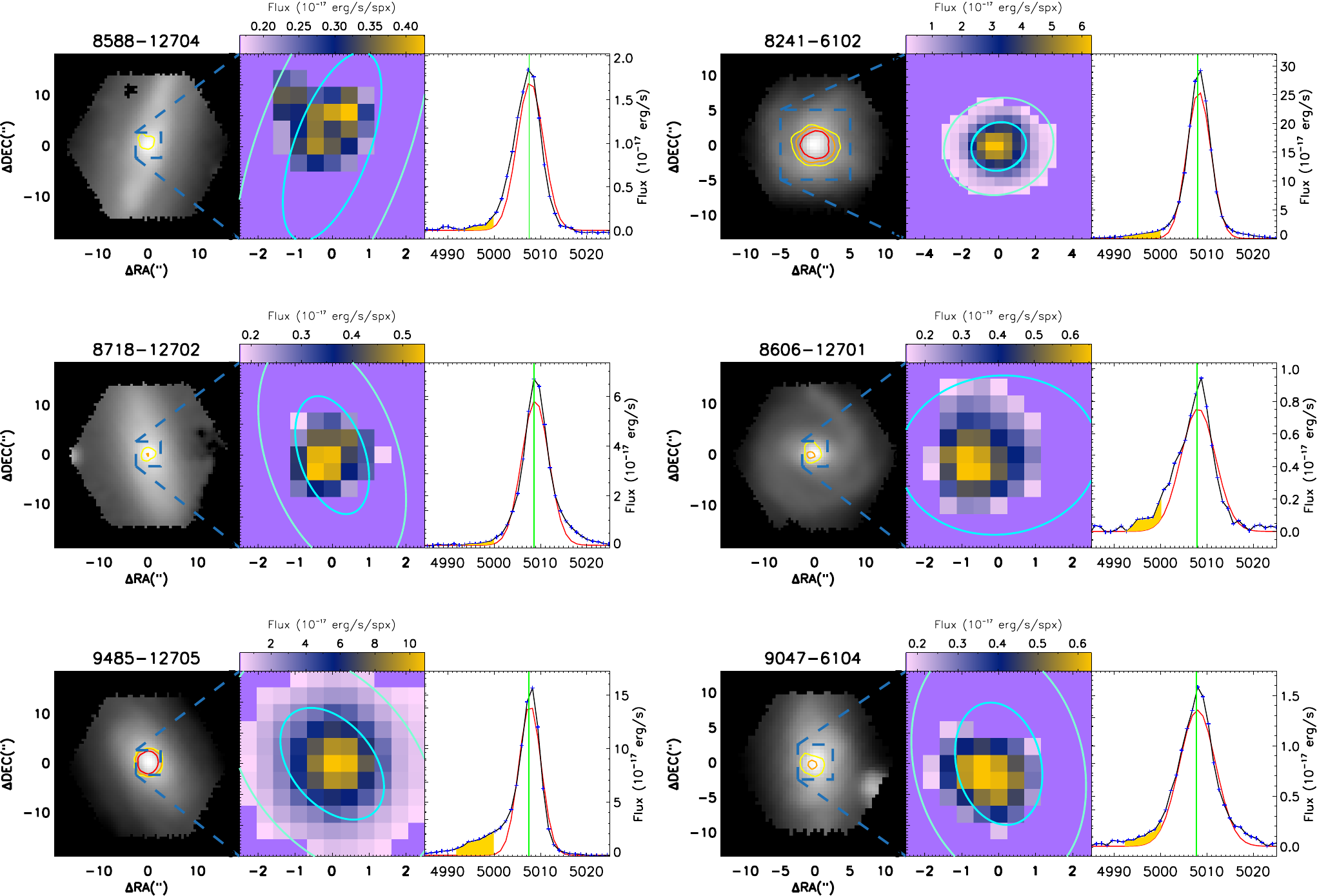}
\caption{ The six AGN with [OIII]5007\AA\ outflows. For each object, from left to right,
    the broad-band image with the distribution of outflow S/N overlaid, where the yellow, orange and red 
    contour correspond to S/N equals to 3, 7.5 and 25, respectively; the map of [OIII]5007\AA\ outflow strength (integrated between -500 and -1000 km/s); 
    the integrated [OIII]5007\AA\ line from the central 1 arc-second where the filled yellow region indicates the outflow. 
    The inner and outer ellipse represents the projected circle with radius of 0.25$R_{e}$ and 0.5$R_{e}$}
    \label{agnsample_outflow}
\end{figure*}

Based on the data of 14 AGNs, our analysis shows a marginal suppression of star formation 
within central few kpc. This result suggests moderate luminosity AGN could regulate 
the star formation in host galaxies. AGN outflows in ionized gas are detected in some of our samples 
by the residual flux at high velocity(in the range of blueshifted by 500-1000km/s) after subtracting the 
gaussian-fitted narrow emission line and stellar continuum from the spectra. 
Fig.~\ref{agnsample_outflow} shows the distribution of outflow signals with S/N>3 per pixel for 6 of our sample. 
Most of these detected 
outflow are confined to the central few arc-seconds where suppressed star formation is identified. The small scale outflows
could probably be more pervading in the sample than we identified, since the spatial resolution and low radial velocity of outflowing gas 
limit the detection rate under this coarse method(see \citet{Wylezalek17} for an example, 8459-3702, which is also included in our sample) 

It is possible that the outflows co-spatial with the low-SFR region cause the star formation suppression. 
However, in some cases the ability of ionized gas outflows to impact the central star formation is questioned, as they 
could only carry limited energy and momentum that
is not likely to significantly influence the dense gas and the star formation therein\citep{Balmaverde16, Bae17}.
In recent studies, more convincing evidence of AGN feedback on star formation has been revealed by the discovery of
growing number of AGNs with outflows that could remove dense molecular 
gas within the central region of the host galaxies\citep{Cicone14, Garcia-Burillo14}, which is more 
closely related to ongoing star formation in galaxies. However, single dish observations
still find similar central molecular gas mass and gas fraction in local AGNs as compared normal 
galaxies whose stellar mass and morphology are controlled to be the same\citep{LLAMA18}, indicating that the outflow can
not significantly impact the dense molecular gas content of the galaxies. This is consistent with our finding that central 
SFRs in AGNs are only mildly suppressed.

Besides the feedback by outflow, moderate to low luminosity AGNs could also impact the star formation by the
kinetic energy carried in small scale jets\citep{Guillard15, Querejeta16}, which represent another probable paradigm of AGN feedback in Seyfert galaxies. 
Enhancement in turbulence caused by the interaction between outflows/jets with ambient gas could provide additional support against the self gravity, suppressing the collapse of gas clouds and star formation.
For our samples, we find not all of our AGNs detected in radio continuum show lower sSFR in their centers. This might be caused by the small physical scale 
of regions affected by jets compare to MaNGA's resolution. The gas distribution in general follow rotational disk of host galaxies, while the orientation of jets are not strongly correlated with the large scale angular momentum of galaxies. As a result the jets
could interact with the dense molecular gas and probably impact the star formation in their host galaxies only when the jets are nearly 
coplanar with the disks and dense ISM is distributed along their path. 

In our study we measured the star formation rate by H$\alpha$ emission \citep{Kennicutt12} that trace short-lived (10 Myrs) O stars. Our optical selection of AGNs also only sensitive to SMBHs undergo active accretion of gas, which generally has a duration of $10^7-10^8$ years. \citep{Haehnelt93, Novak11}. Therefore our result reflects the impact of current AGN activity on the recent star formation in host galaxies. Literatures usually refers to 
this kind of feedback coherent with AGN activity as fast mode of feedback\citep{Alexander12, Harrison18, Cresci18}, where the radiation/outflow/jet released during 
the accretion of SMBHs directly impact the star formation. Fast mode feedback has been suggested to be responsible for the fast star
 formation quenching in major mergers triggering luminous quasars\citep{Springel05, Hopkins06}, which probably result in the massive non-active galaxies at higher redshift. 
 Our results suggest that it probably also have mild impact on the star formation in local disk galaxies.

A growing number of studies supported a generally long quenching timescale in local quiescent galaxies\citep{Schawinski14, Peng15, Belfiore17_gv, Sanchez18}. These studies pointed out that it is the gradual exhaustion of internal gas or(and) the secular growth of central bulges that contributes most to the galaxy quenching in local universe, which acts at timescales of around 1 Gyrs. Studies on the integrated colors of late type galaxies indicated that AGNs in these galaxies are in favor of more aged stellar population than galaxies that have just started quenching, which does not support a casual link between global quenching of late type galaxies and nuclear activities\citep{Schawinski10, Schawinski14}. However, it might be difficult for previous studies to search for spatially limited AGN feedback on star formation using integrated physical qualities of galaxies. Our discovery of marginally suppressed star formation within central kpc of MaNGA AGNs provide evidence that the fast removal/exhaustion of gas by AGN feedback could also act as an important process in the quenching of central part of secular evolved galaxies, in addition to the slower quenching process that has been widely suggested. 


Alternative explanations to our finding such as the AGNs in favor of more aged stellar populations exceeding the lifetime of massive hot stars could also be plausible, which has been suggested by previous study like \citet{Norman88, Davies07}. They suggest the slower stellar wind from aged stars could provide material more easily accreted by central black holes accretion. Recent studies has also revealed qualitatively similar overabundance of stellar population within the central 0.5 Re of some more luminous MaNGA AGNs\citep{Rembold17,Mallmann18}. However, comparing to our studies, these works either focus on stellar population at much smaller physical scales, or use a different controlling method in their comparison. In order to reveal if these results and our finding reflect different sides of interaction between AGN and central star formation in host galaxies, more detailed studies on the stellar population of current sample are needed.

\section{Conclusion}

We investigated the spatially-resolved star formation 
in 14 local Seyfert galaxies with the IFU observations by the  SDSS-IV MaNGA survey. For each of 14 AGNs, we carefully selected a set of 
normal galaxies with the total stellar mass, bulge to disk ratio, central stellar mass surface density and specific star formation rate at 1.0-1.5$R_{e}$ 
controlled to be the same.We derive the specific star formation rates of 14 Seyfert galaxies within 5 radial bins measured from IFU spectra and compared with those of the control sample. 
The comparison shows that the central radial bins of AGN have slightly lower SFRs with false-positive possibilities of 0.76\% in the innermost 0.3Re and 3.5\% within the 
0.3-0.6Re. These low possibilities indicate marginally suppressed star formation within the central region of our AGNs. This may suggest that moderate SMBH accretion is capable of regulating star formation at the galaxy centers.

\section*{Acknowledgements}

The authors thank Yifei Jin, Christy Tremonti and Renbin Yan for their valuable suggestions.
L.B. and Y.S. acknowledge support from the National Key R\&D Program of China 
(No. 2018YFA0404502), the National Natural Science Foundation of China 
(NSFC grants 11733002 and 11773013), the Excellent Youth Foundation of the Jiangsu 
Scientific Committee (BK20150014), and National Key R\&D Program of China 
(No. 2017YFA0402704). Y.C. acknowledge support from the National Natural Science 
Foundation of China (NSFC grants 11573013). SFS thank the CONACyT 
program CB-180125 and DGAPA-PAPIIT IA101217 grants for their support 
to this project. R.M. acknowledges support by the Science and Technology 
Facilities Council (STFC) and from the ERC Advanced Grant 695671 "QUENCH".
RR thanks to CNPq and FAPERGS. D.B. is supported by grant RScF 14-50-00043. 

Funding for the Sloan Digital Sky Survey IV has been provided by the Alfred 
P. Sloan Foundation, the U.S. Department of Energy Office of Science, and the 
Participating Institutions. SDSS-IV acknowledges
support and resources from the Center for High-Performance Computing at
the University of Utah. The SDSS web site is www.sdss.org.

SDSS-IV is managed by the Astrophysical Research Consortium for the 
Participating Institutions of the SDSS Collaboration including the 
Brazilian Participation Group, the Carnegie Institution for Science, 
Carnegie Mellon University, the Chilean Participation Group, the French 
Participation Group, Harvard-Smithsonian Center for Astrophysics, 
Instituto de Astrof\'isica de Canarias, The Johns Hopkins University, 
Kavli Institute for the Physics and Mathematics of the Universe (IPMU) / 
University of Tokyo, Lawrence Berkeley National Laboratory, 
Leibniz Institut f\"ur Astrophysik Potsdam (AIP),  
Max-Planck-Institut f\"ur Astronomie (MPIA Heidelberg), 
Max-Planck-Institut f\"ur Astrophysik (MPA Garching), 
Max-Planck-Institut f\"ur Extraterrestrische Physik (MPE), 
National Astronomical Observatories of China, New Mexico State University, 
New York University, University of Notre Dame, 
Observat\'ario Nacional / MCTI, The Ohio State University, 
Pennsylvania State University, Shanghai Astronomical Observatory, 
United Kingdom Participation Group,
Universidad Nacional Aut\'onoma de M\'exico, University of Arizona, 
University of Colorado Boulder, University of Oxford, University of Portsmouth, 
University of Utah, University of Virginia, University of Washington, University of Wisconsin, 
Vanderbilt University, and Yale University.

This project makes use of the MaNGA-Pipe3D dataproducts. We thank the IA-UNAM 
MaNGA team for creating this catalogue, and the ConaCyt-180125 project for supporting them.

This research made use of Marvin, a core Python package and web framework 
for MaNGA data, developed by Brian Cherinka, Jos\'e S\'anchez-Gallego, and 
Brett Andrews. (MaNGA Collaboration, 2017).




\bibliographystyle{mnras}
\bibliography{citation}  








\bsp	
\label{lastpage}
\end{document}